\documentclass{du-journals}
\journal{Iranian Journal of Astronomy and Astrophysics}
\pdfoutput=1
\title{The Main Sources of Gas Ionization in Several Nearby Star-Forming Galaxies }
\subtitle{The main ionization sources of DIG}
\author[1]{Behjat Zarei Jalalabadi}
\address[1]{Department of Physics, University of Birjand, Birjand, Iran; \\ email: zarei.b@birjand.ac.ir}

\author[2]{Abbas Abedi}
\address[2]{Department of Physics, University of Birjand, Birjand, Iran ;\\ email:  aabedi@birjand.ac.ir}

\author[3]{Alexei V.  Moiseev}
\address[3]{Special Astrophysical Observatory, Russian Academy of Sciences, Nizhnij Arkhyz, 369167, Russia;\\Space Research Institute, Russian Academy of Sciences, Moscow, 117997 Russia;\\ email:  moisav@gmail.com}
\begin{document}
\begin{abstract}
The emission-line intensity ratios are used to distinguish the main sources of gas ionization to study the state of galactic interstellar medium (ISM). In intermediate cases, when the contributions of radiation from OB stars and from shock waves mix, the identification becomes uncertain. As an extra parameter, the gas velocity dispersion in the line-of-sight can be added to classical diagnostic diagrams (i.e., "BPT-$\sigma$" relations) to help finding an appropriate solution. The minimum distance from the curve that bounds the H II-type ionization region for each point in BPT-$\rho$ diagram can be used to characterize the excitation mechanism of the ionized gas. The shock excitation in the diffuse ionized gas (DIG) can be realized by the correlation between $ \rho $ and $  \sigma$, while the H II regions with low level turbulent motions can be characterized by the absence of this correlation. We consider the "BPT-$\sigma$" relation and the correlation between $\sigma$ and $\rho$ to determine the ionized gas excitation in several nearby star-forming galaxies. Distributions of the velocity dispersion are obtained from the scanning Fabry-Perot interferometer observations at the SAO RAS 6-m telescope, whereas the emission-line ratios are calculated from the archival long-slit spectroscopic data. The results of this study are reported for Mrk 370, NGC 4068, UGC 8313, and UGC 8508.
\end{abstract}


\section{Introduction}

To study the state of galactic interstellar medium (ISM), the emission-line intensity ratios are used to distinguish the main ionization sources, such as hot massive young OB stars in the star formation H II regions, active galactic nuclei (AGNs), shock waves related to supernova remnants (SNR), and other feedback processes like supersonic turbulence generated by stellar winds. In intermediate cases, for example, when the contributions of radiation from OB stars and shock waves mix, the identification becomes challenging, and the issue remains unresolved on what determines the observed conditions of the diffuse ionized gas (DIG) including the one on large distances from the galactic plane.As an extra parameter, the gas velocity dispersion in the line-of-sight can be added to classical diagnostic diagrams to help finding an appropriate solution. 
Baldwin, Phillips \& Terlevich (1981)[1] proposed a diagram by comparing the line intensity ratios [O III]$\lambda$5007/H$\beta$ with [N II]$\lambda$6583/H$\alpha$  to separate the emission from soft-ionizing sources, like H II regions, and objects with a higher ionizing power such as AGNs. Subsequently, other researchers extended this scheme incorporating diagnostic procedures. All of the mentioned diagrams are frequently referred to as "BPT diagrams" in the literature after the authors of the method. One of the most effective methods for studying the extended low-brightness structures in galaxies is 3D (or integral-field) spectroscopy. Two surveys on 3D spectroscopy were presented by MaNGA [2] and SAMI [3]. A significant limitation of these two comprehensive 3D spectroscopy surveys of galaxies is their low spatial resolution (more than 1 kpc). This limitation is caused by a relatively small field of view (FOV) of these instruments. We collect spectral data simultaneously with a relatively high spectral resolution (to resolve the emission line width) and spatial resolution better than 100-300 pc (a few arcsec in the observed line-of-sight). Velocity dispersion, which is a characteristic of turbulent motions in an ionized gas, can occur due to several causes. Various factors affect line flux relations with different excitation mechanisms, so we expand a sample of objects to study the "BPT-$\sigma  $" relation in the ISM of galaxies.

\section{Observations and Data Reduction}

Our data were extracted from about 10-20 galaxies using the SAO RAS 6-m telescope (BTA) data archive (ASPID)[4] observed with the multi-mode focal reducers SCORPIO and its improved version SCORPIO-2 at the prime focus (Afanasiev $ \& $ Moiseev, 2005, 2011)[5,6]. The main samples are spiral and dwarf galaxies that were observed in the scanning Fabry–Perot interferometer (FPI) SCORPIO/SCORPIO-2 mode with a high signal-to-noise ratio of the ionized gas velocity dispersion maps. For each galaxy in the list, we checked the presence of data taken from the long-slit spectrograph mode in a wide optical range. Spectral data were obtained using the SCORPIO multi-mode spectrograph (Afanasiev $\&$ Moiseev 2005)[5]. Our Data are collected between 2012- 2015. We used a VPHG550G grism covering the range of 3500-7500 A° with a 1.0 arcsec slit width, which provides a typical spectral resolution of 13 A° as estimated from the full-width-half-max (FWHM)of air-glow emission lines. Finally, we combine the ionized gas velocity dispersion maps derived from the scanning FPI observations at the SAO RAS 6-m telescope with the emission line ratios obtained from the archival long-slit spectroscopic data. The archival observational data are reduced using an Interactive Data Language (IDL)-based software developed for SCORPIO/SCORPIO-2 data reduction in the laboratory of spectroscopy and photometry of extragalactic objects in SAO RAS (the Special Astrophysical Observatory of the Russian Academy of Sciences)[7].

\section{Our Samples}

\begin{table*}
\begin{center}
\caption{ Characteristics of our samples and corresponding observational parameters  
(SAO RAS 6-m telescope data [4])\label{table1}
}
\begin{tabular}{cccccccc}
\hline
Galaxy & D(Mpc)    &   $M_{B}$ & Instrument & Spectroscopy   & & Scanning FPI  &   \\
& & & &$\Delta \lambda (A^{\circ})$ & $\delta \lambda (A^{\circ})$ & $\Delta \lambda (A^{\circ})$  & $\delta \lambda (A^{\circ})$     \\

\hline
&&&&&&&\\

Mrk 370 & 10.85 & -16.83 &SCORPIO & 3500-7500 &13 & H$ \alpha $ & 0.8      \\
NGC 4068 & 4027 & -15.17 & SCORPIO-2 &3500-7500 &13 & H$ \alpha $ & 0.4   \\
UGC 8313 & 9.65 & -14.77 &SCORPIO-2& 3500-7500 &13 & H$ \alpha $ & 0.8    \\
UGC 8508 & 2.69 & -13.09 &SCORPIO/SCORPIO-2& 3500-7500 &13 & H$ \alpha $ & 0.8   \\
  \\\hline

\end{tabular}
\end{center}
\end{table*}
\textbf{Mrk 370} belongs to Blue Compact Dwarf (BCD) galaxies. BCDs are low luminosity $(M_{B} \geq -18$ mag) and compact (starburst diameter $  \leq$ 1 kpc) galaxies. Mrk 370 has a chain morphology. Moiseev (2010) [8] has considered the presence of polar gaseous structures in Mrk 370 due to current burst of star formation and external gas accretion or merging.

\textbf{NGC 4068} is a BCD-type galaxy. Bastian et al. (2011) [9] found 60 OB stars in NGC 4068 by analyzing stellar population.

\textbf{UGC 8313} is a companion galaxy of NGC 5055 and a small SBc-type galaxy (Tully (1988) [10].

\textbf{UGC 8508} is a irregular dwarf galaxy and Kraan-Korteweg $ \&$ Tammann(1979) [11] listed it as a possible  member of the M 101 group.

\section{BPT-$  \sigma$ and $  \rho$-$  \sigma $ Diagrams}

We define the velocity dispersion of ionized gas ($  \sigma$) as the standard deviation of the Gaussian profile fitted to the H$  \alpha$ emission line after accounting for the FPI instrumental profile and subtracting the contribution of thermal broadening in the H II regions. The procedure to measure $  \sigma$ is described in detail by Moiseev $ \&$ Egorov(2008)[12].
In short, the observed profiles of the H$  \alpha$ line were fitted by the Voigt function, which is a convolution of Lorentzian and Gaussian functions corresponding to the FPI instrumental profile and broadening of observed emission lines, respectively. The FWHM of the instrumental profile was estimated each night from Lorentzian fitting of the He-Ne-Ar calibration lamp emission scanned using FPI. The results of the profile fitting were used to construct two-dimensional line-of-sight velocity fields of ionized gas, maps of line-of-sight velocity dispersion free from the instrumental profile influence, as well as the images of galaxies in the H$  \alpha$ emission line and in the continuum (Egorov et.al 2018)[13]. 
 
According to Kewley et al. (2001, 2006)[13,14], the minimum distance from the curve that bounds the H II ionization region for each point on the BPT diagram (represented by $  \rho$) can be important in characterizing the excitation mechanism of the ionized gas (Oparin $ \&$ Moiseev (2019)) [14]. The negative values of $ \rho $ correspond to the points that are obtained from photoionization caused by young stars, while the positive values correspond to the data from other ionization mechanisms.

We indicated the value of $  \rho$ for the $[N II]/H_{\alpha}$ - $[O III]/H_{\beta}$ diagrams as $\rho (N II)$ and the $[S II]/ H_{\alpha}$ - $[O III]/ H_{\beta}$ diagrams as $  \rho$ (S II).
The shock excitation in DIG can be realized by the correlation between $  \rho$ and $  \sigma$, while the H II regions with low level turbulent motions can be characterized by the absence of correlation between $  \rho$ and $  \sigma$.

\section{Results}

The BPT-$  \sigma$ and $  \rho$-$  \sigma$ diagrams for our samples are shown in Figs. 1-8. The BPT-$  \sigma$ diagrams are divided into different regions according to Kewley et al. (2001, 2006) [13,14].

The red and orange points in Fig. 1 for Mrk 370 with high velocity dispersion are interesting to study. These points indicate a non-stellar mechanism for gas ionization that is confirmed by the correlation between $  \rho$ and $  \sigma$ (Fig. 2).
 
There are some points out of the star formation region in Fig. 3 corresponding to NGC 4068 although the correlation between $  \rho$ and $  \sigma$ in this sample is not observed (Fig. 4).

For UGC 8313, as shown in Fig. 5, all of the points are located in the star formation region. There is a correlation between $  \rho$ and $  \sigma$ for $ \sigma $ $\leq$ 25 km/s (Fig. 6).

Figure 7 shows the results for UGC 8508. Our data is not enough to comment, but it seems that there is not BPT-$  \sigma$ relation in this galaxy (Fig. 8). The BPT diagrams show that all regions with spectral data correspond to the gas ionized by young OB stars.

\begin{figure} 
 \centerline{\includegraphics[width=8cm]{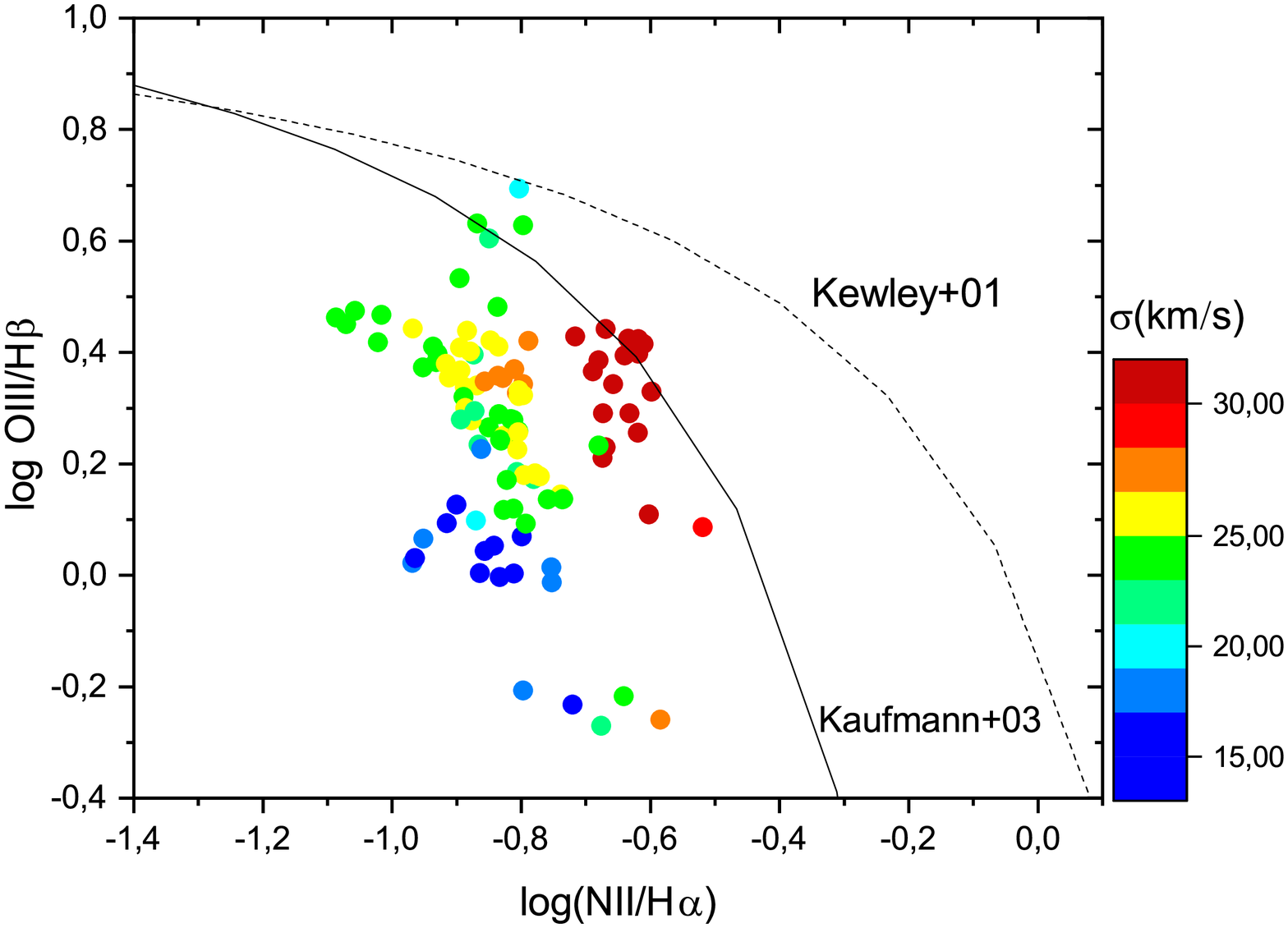}
  \includegraphics[width=8cm]{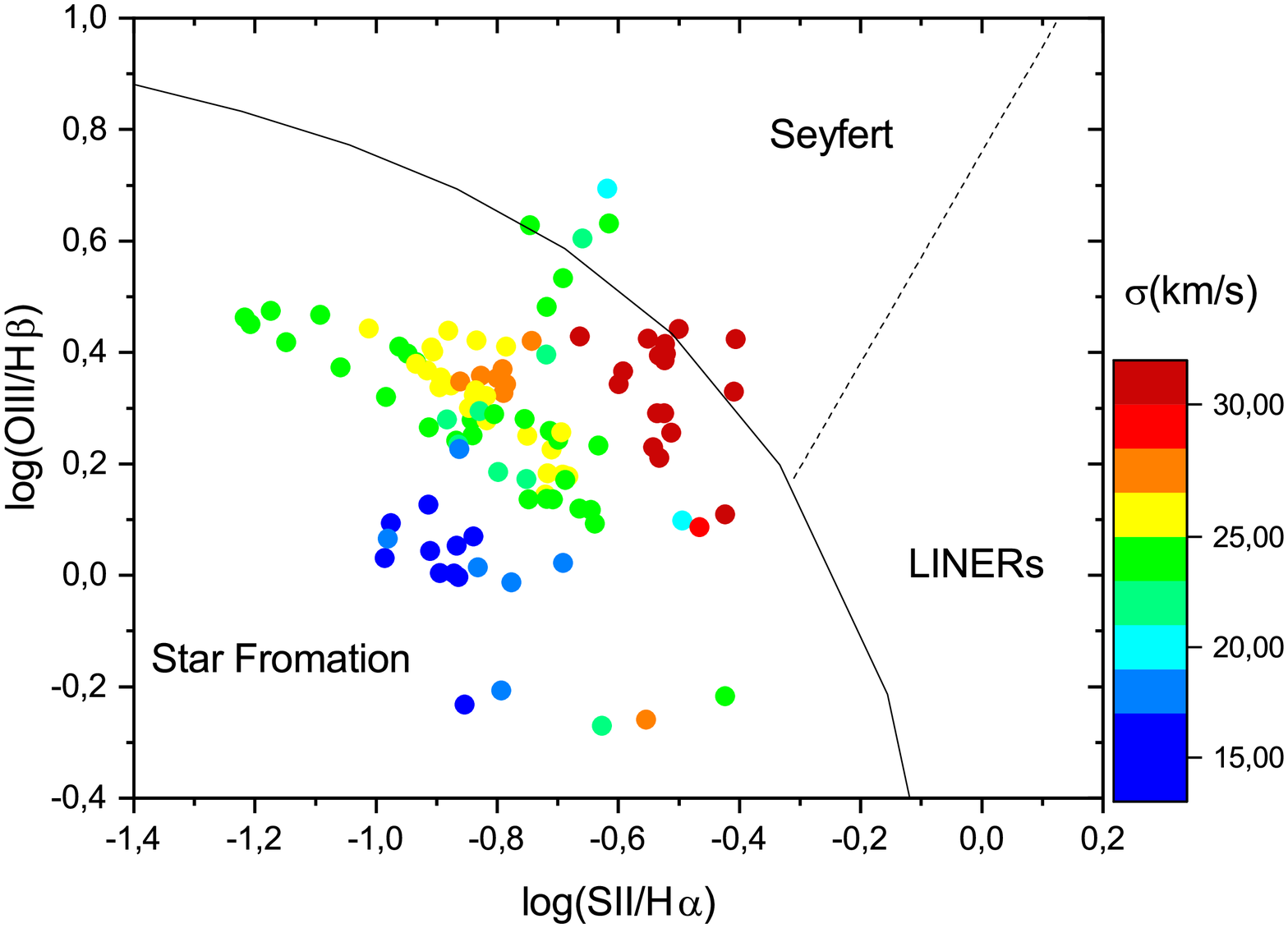}
   }
 \caption[]{Excitation diagnostic diagrams (BPT-$  \sigma$) by comparing the emission-line intensity ratios for Mrk 370 according to SCORPIO-2 long-slit observations. The colored points correspond to different velocity dispersions according to the scale-box.}
 \label{fig1}
\end{figure}

\begin{figure}[h]
\centerline{
\includegraphics[width=8 cm]{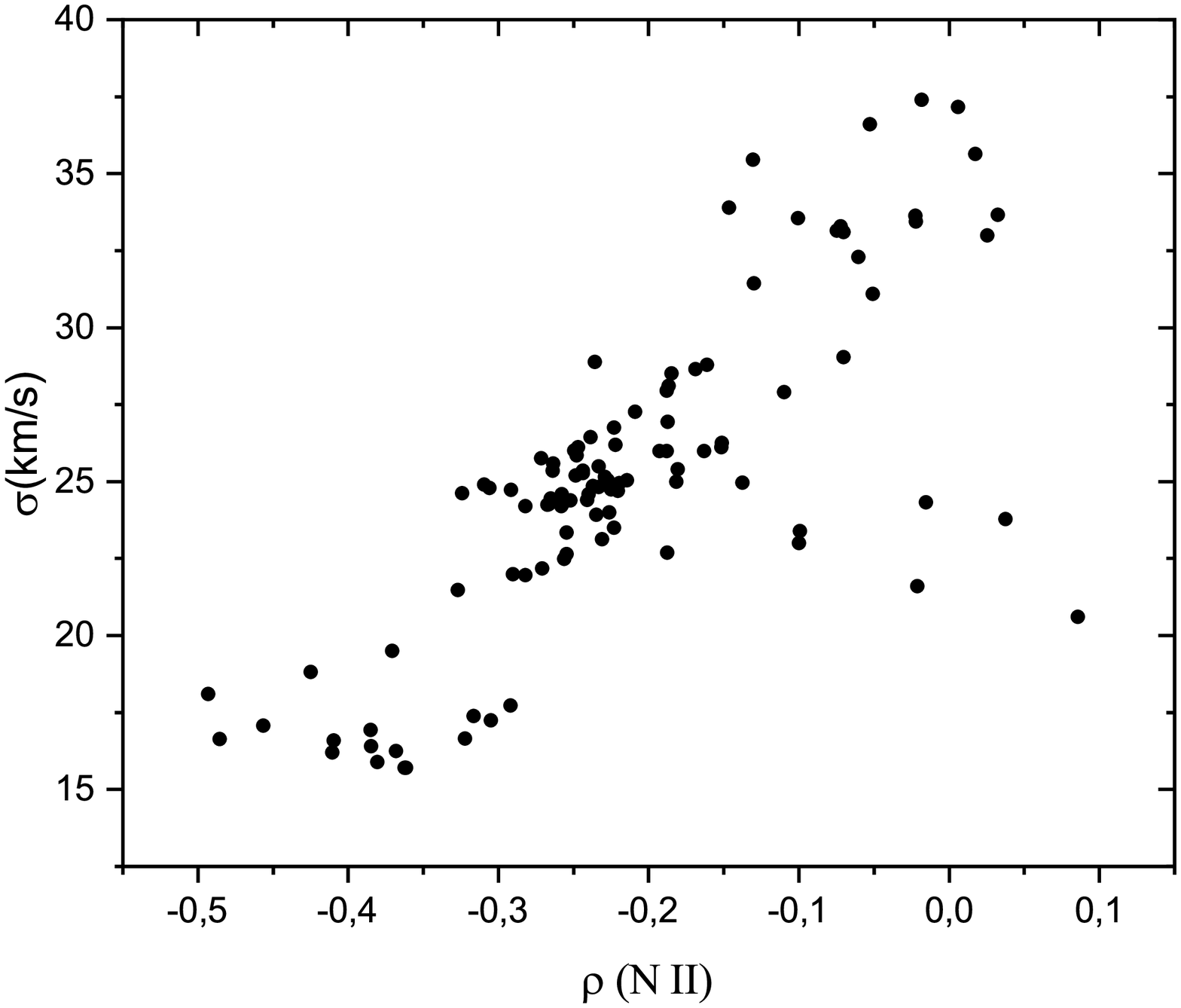}
\includegraphics[width=8 cm]{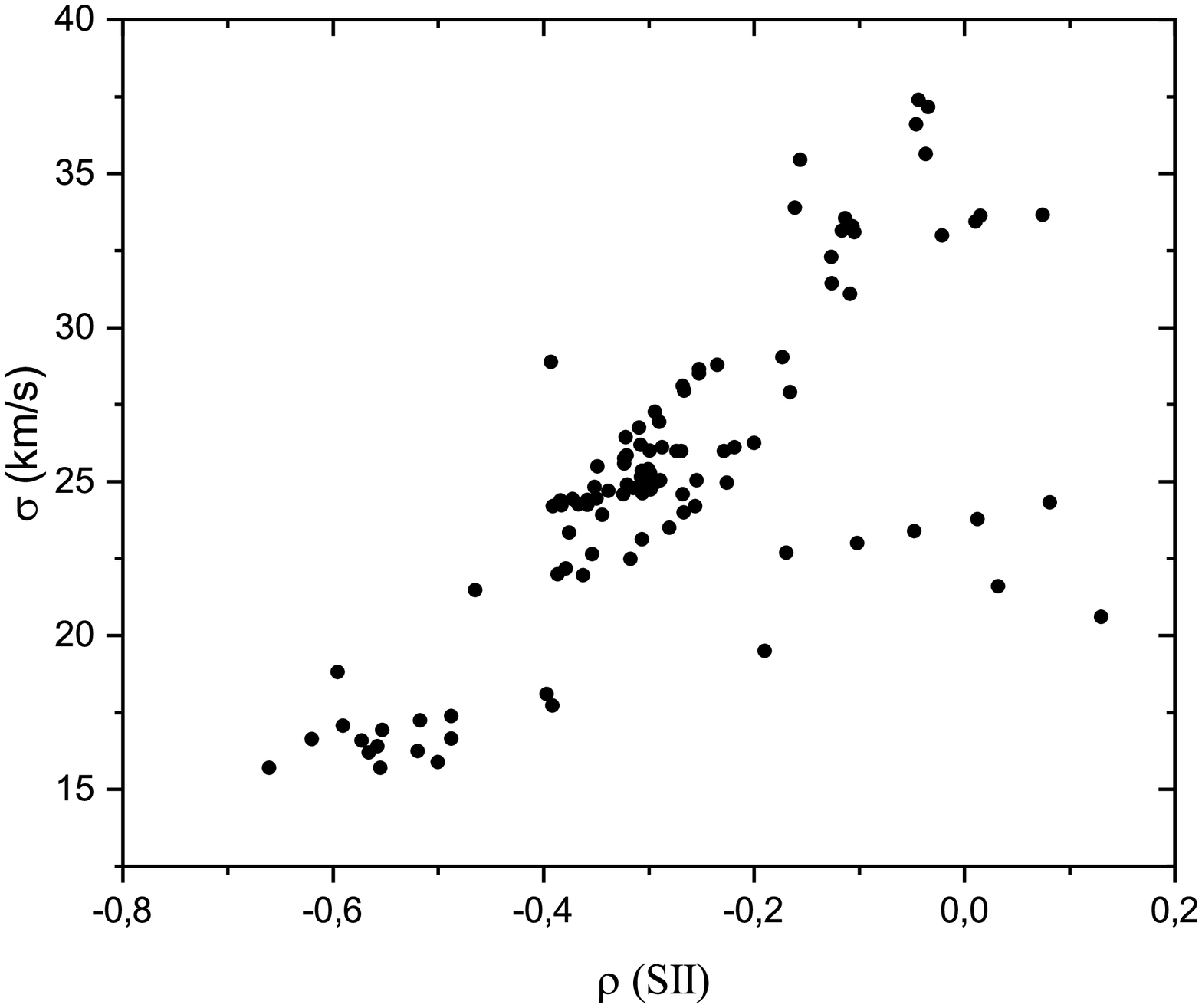} 
}
\caption{Dependence of $\sigma$ on the distance of the point from the demarcation curve in the BPT diagram, which separates the
H II regions and regions with other ionization mechanisms.}
\centering

\end{figure}

\begin{figure} 
 \centerline{\includegraphics[width=8cm]{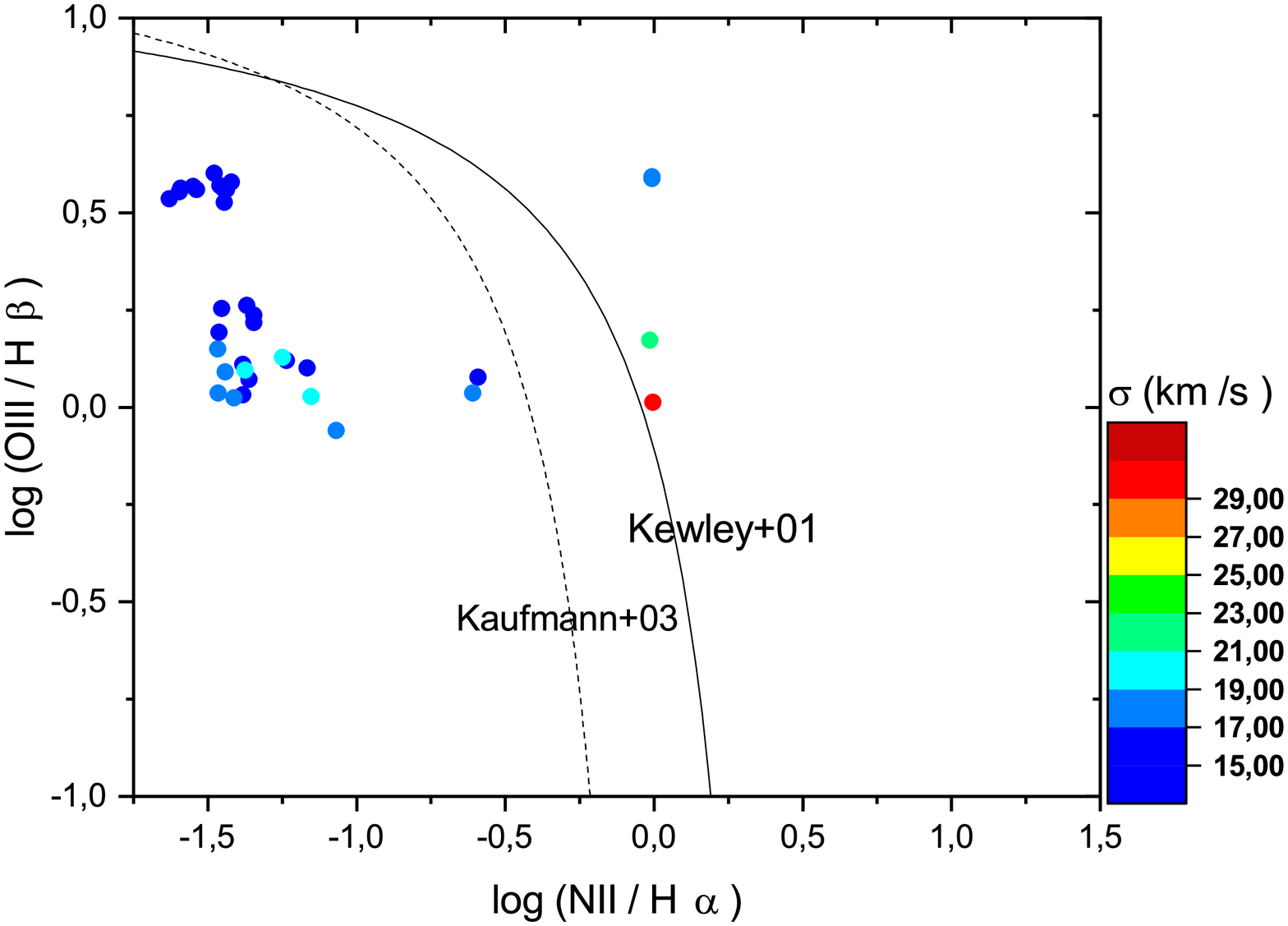}
  \includegraphics[width=8cm]{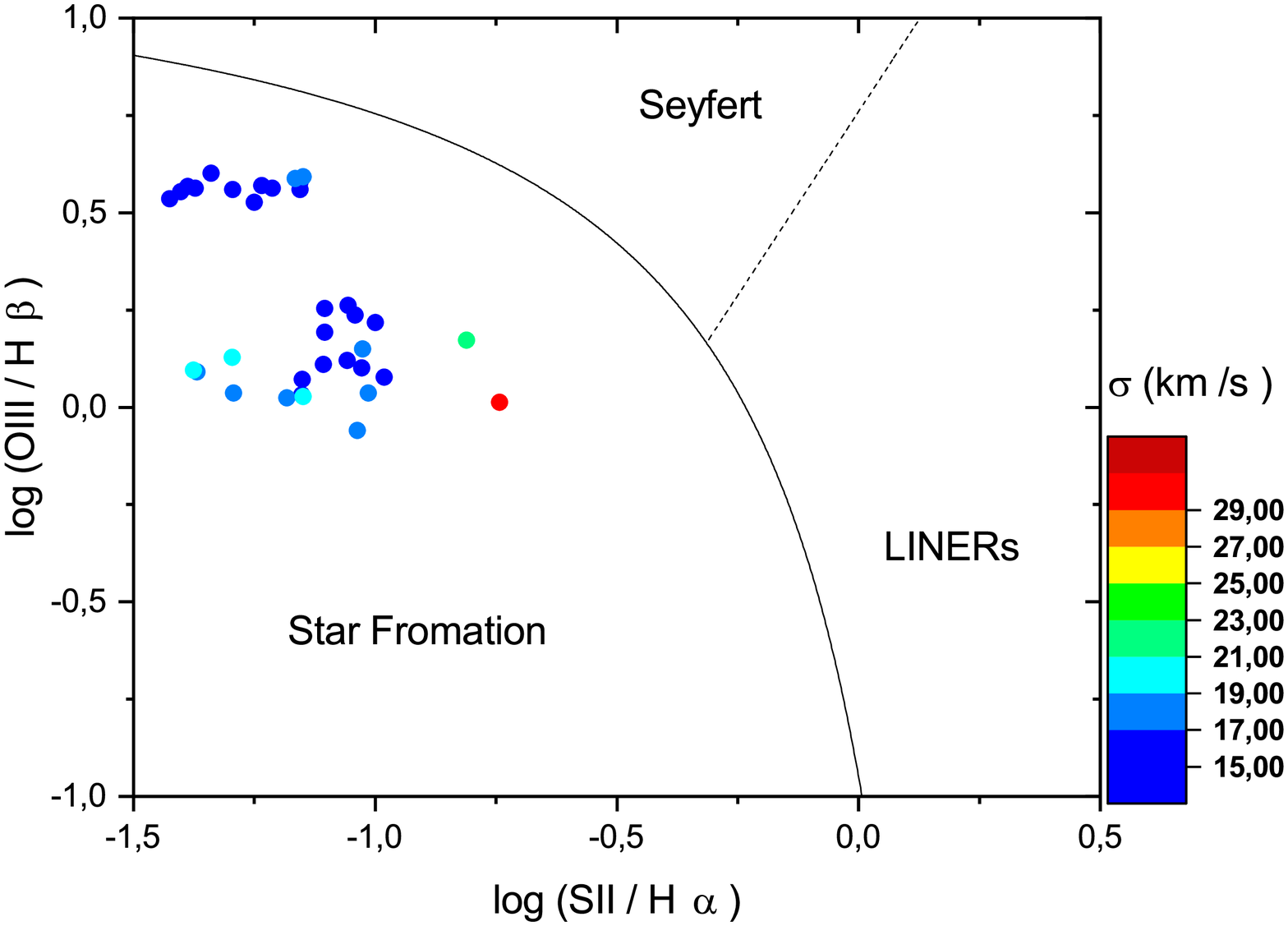}
   }
 \caption[]{The same as Fig. 1 for NGC 4068.}
 \label{fig1}
\end{figure}

\begin{figure}[h]
\centerline{
\includegraphics[width=8 cm]{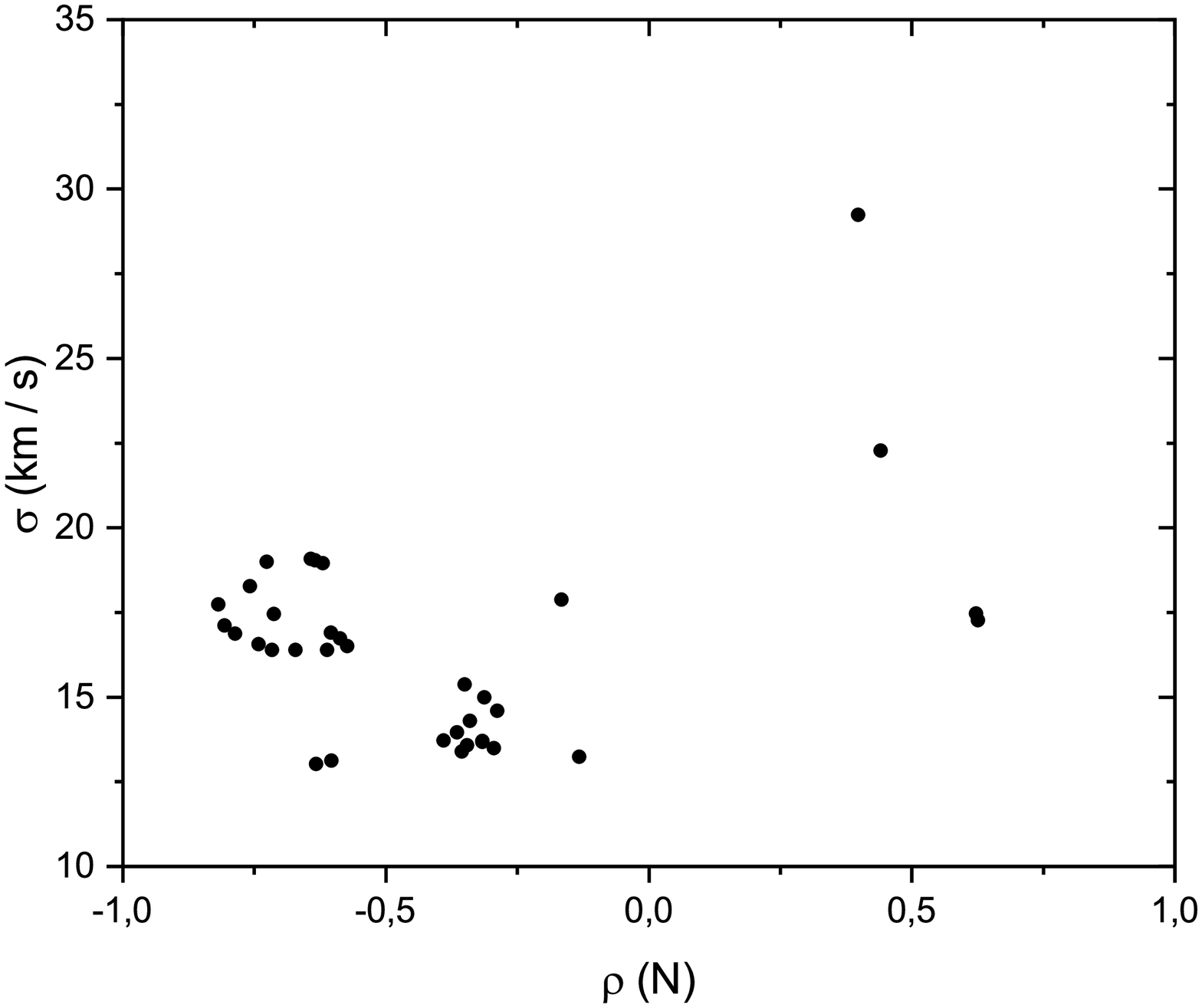}
\includegraphics[width=8 cm]{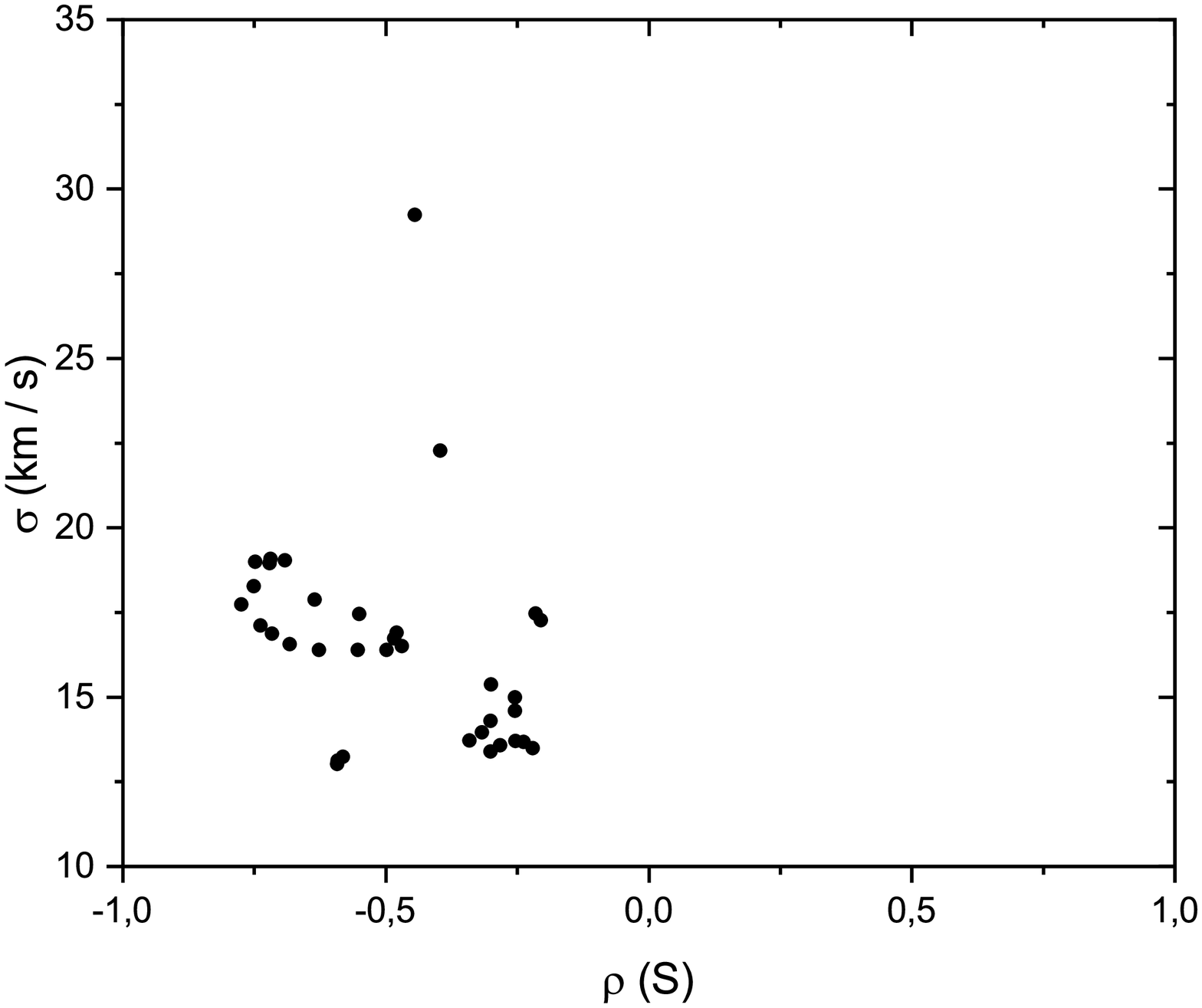} 
}
\caption{ The same as Fig. 2 for NGC 4068}
\centering

\end{figure}

\begin{figure} 
 \centerline{\includegraphics[width=8cm]{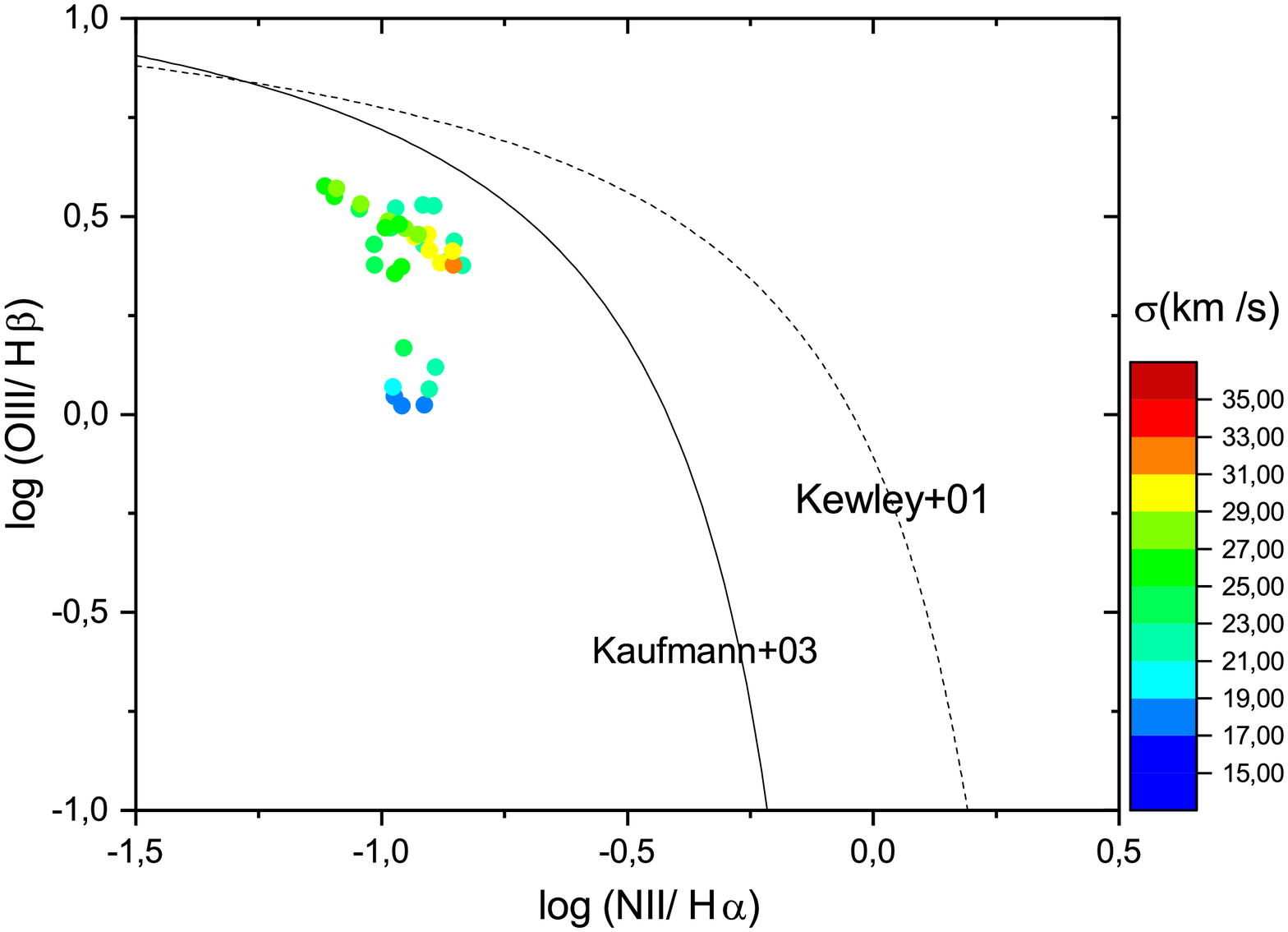}
  \includegraphics[width=8cm]{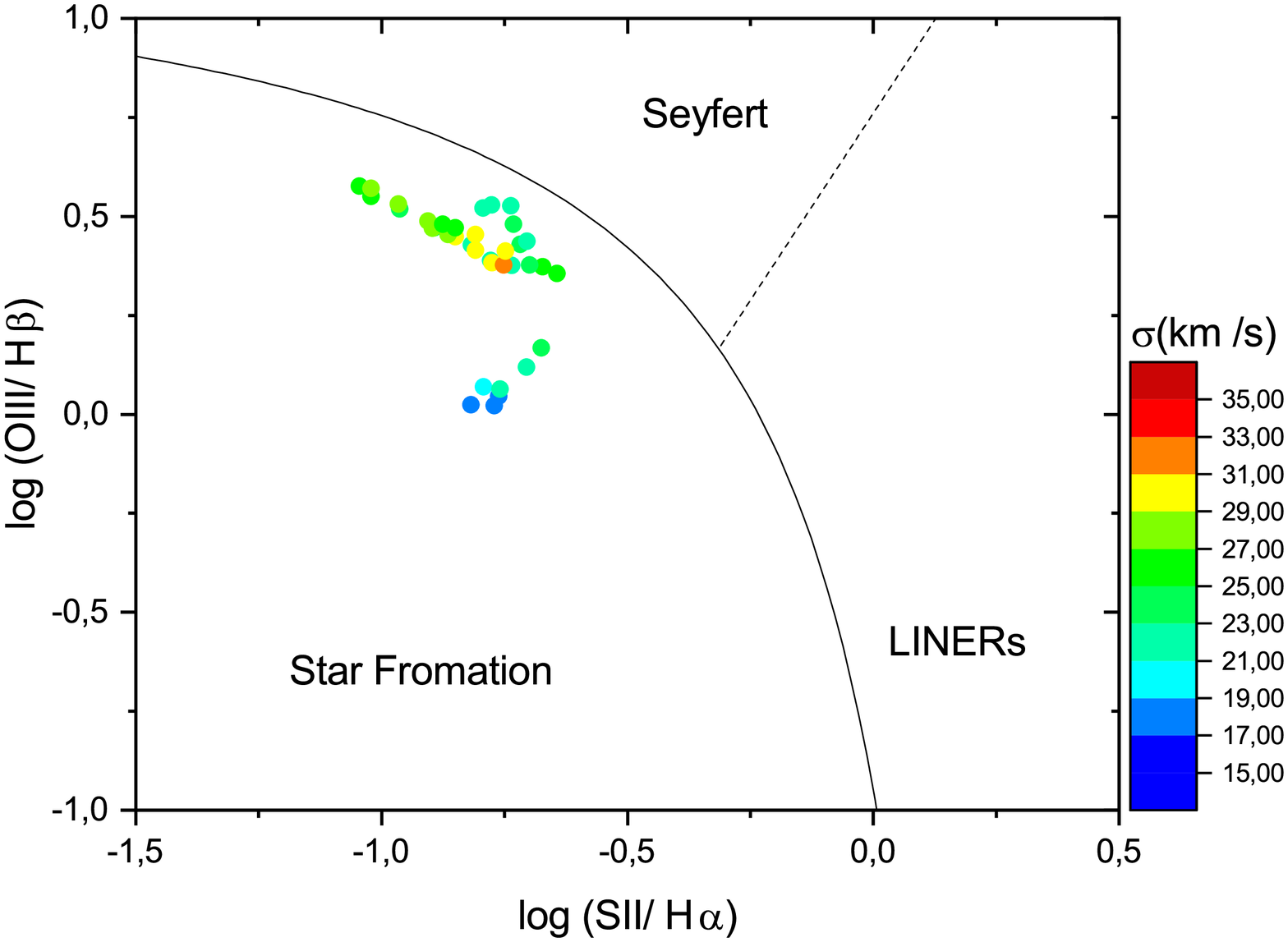}
   }
 \caption[]{The same as Fig. 1 for UGC 8313.}
 \label{fig1}
\end{figure}

\begin{figure}[h]
\centerline{
\includegraphics[width=8 cm]{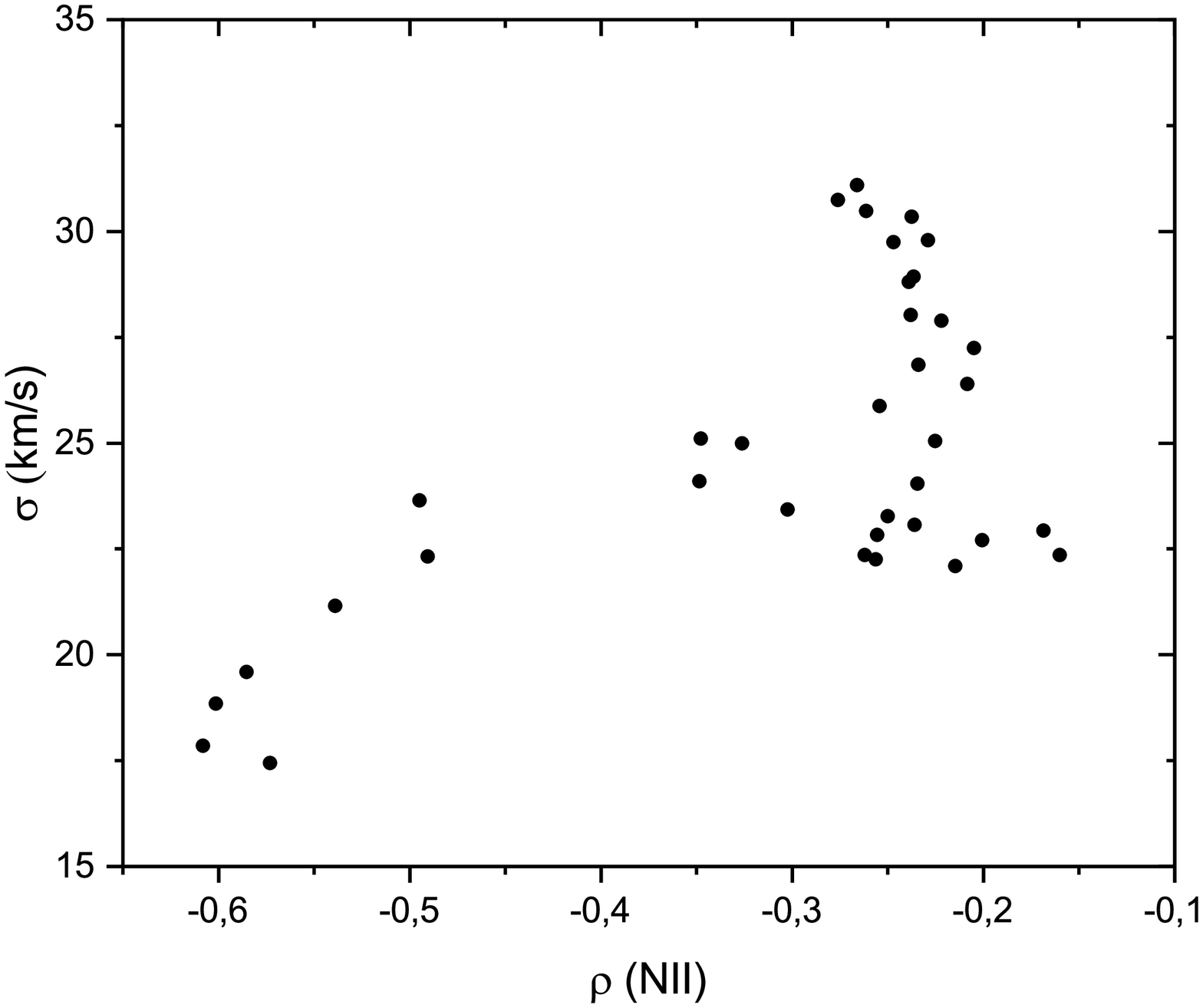}
\includegraphics[width=8 cm]{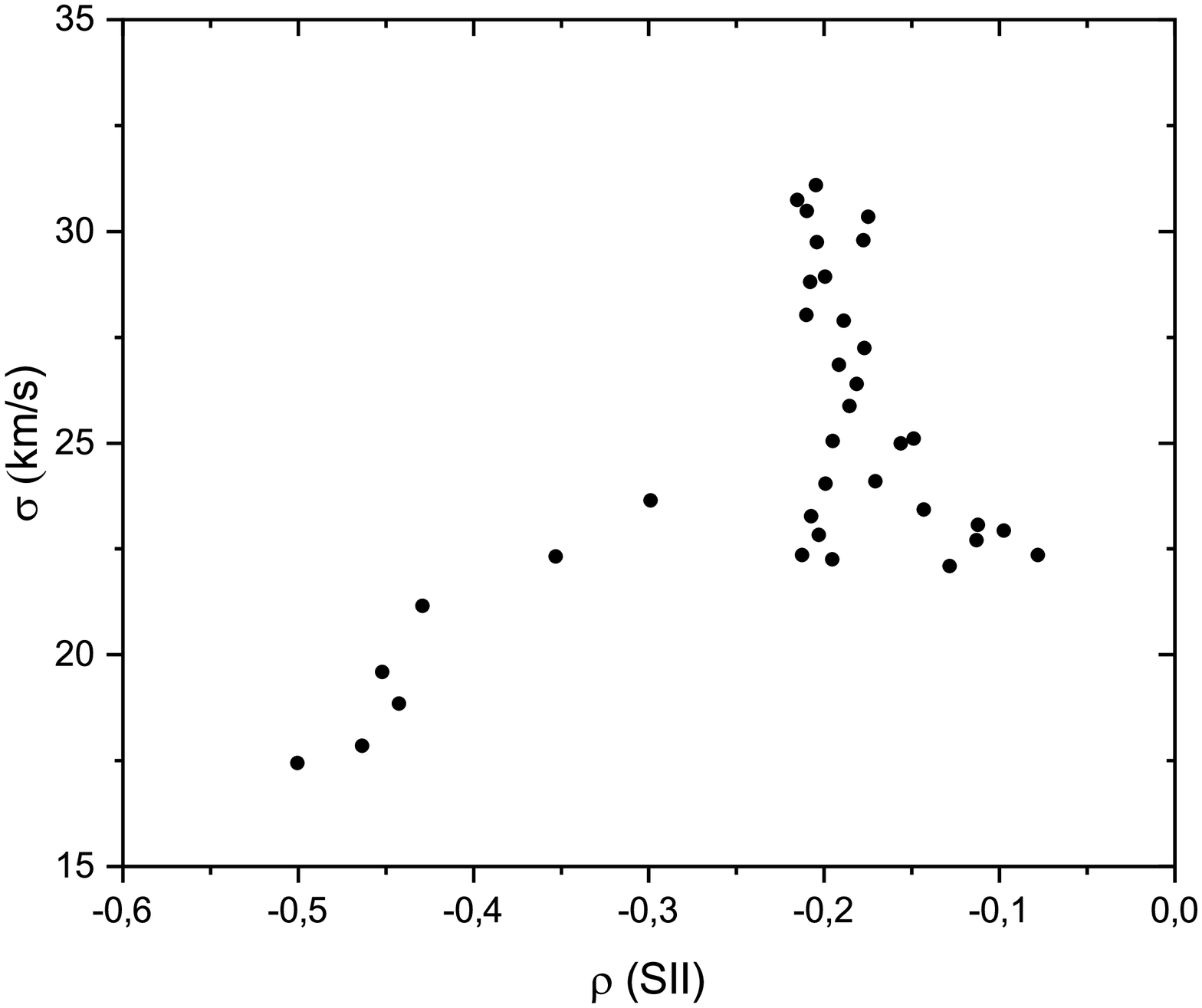} 
}
\caption{ The same as Fig. 2 for UGC 8313.}
\centering

\end{figure}

\begin{figure} 
 \centerline{\includegraphics[width=8cm]{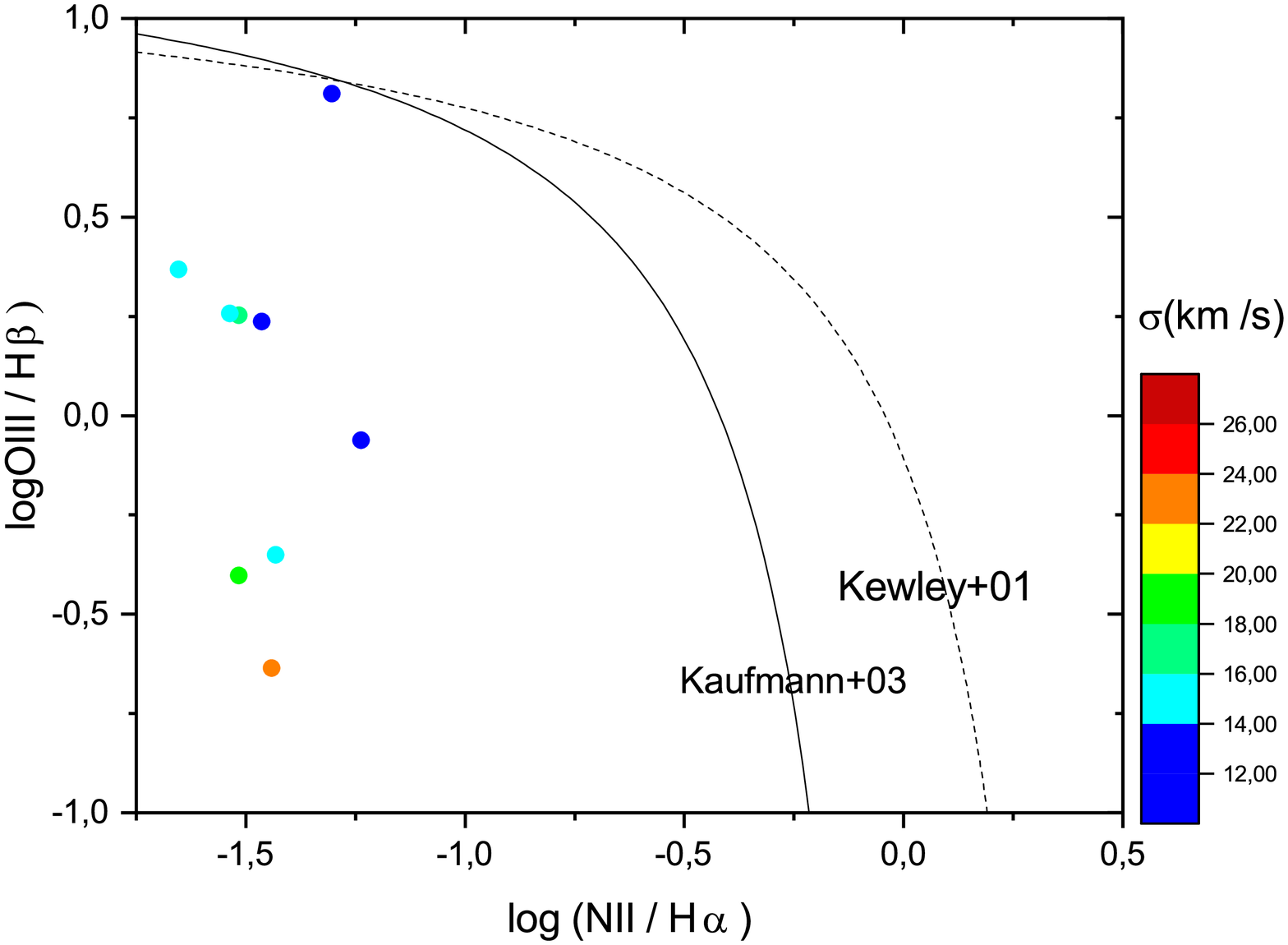}
  \includegraphics[width=8cm]{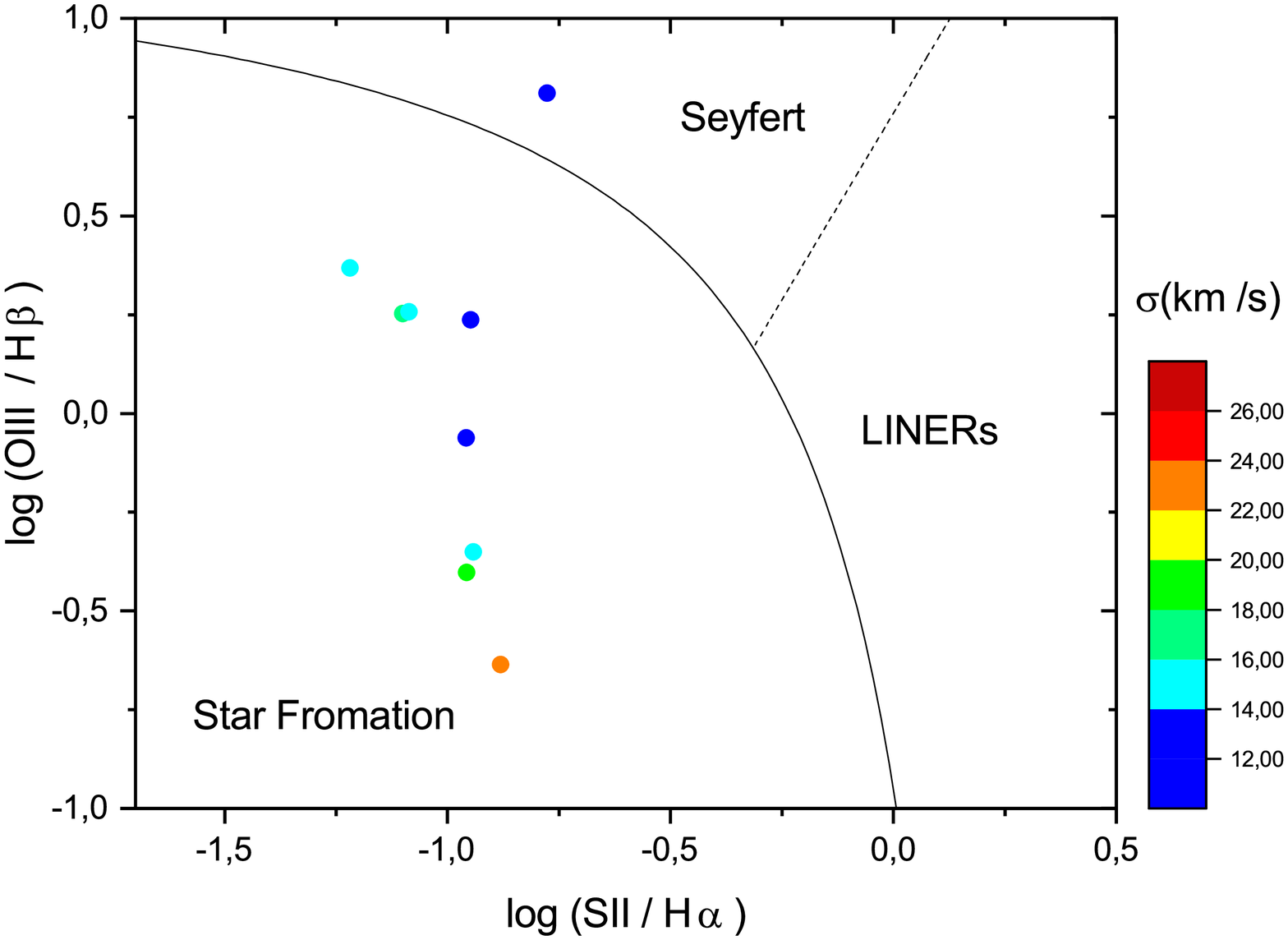}
   }
 \caption[]{The same as Fig. 1 for UGC 8508.}
 \label{fig1}
\end{figure}

\begin{figure}[h]
\centerline{
\includegraphics[width=8 cm]{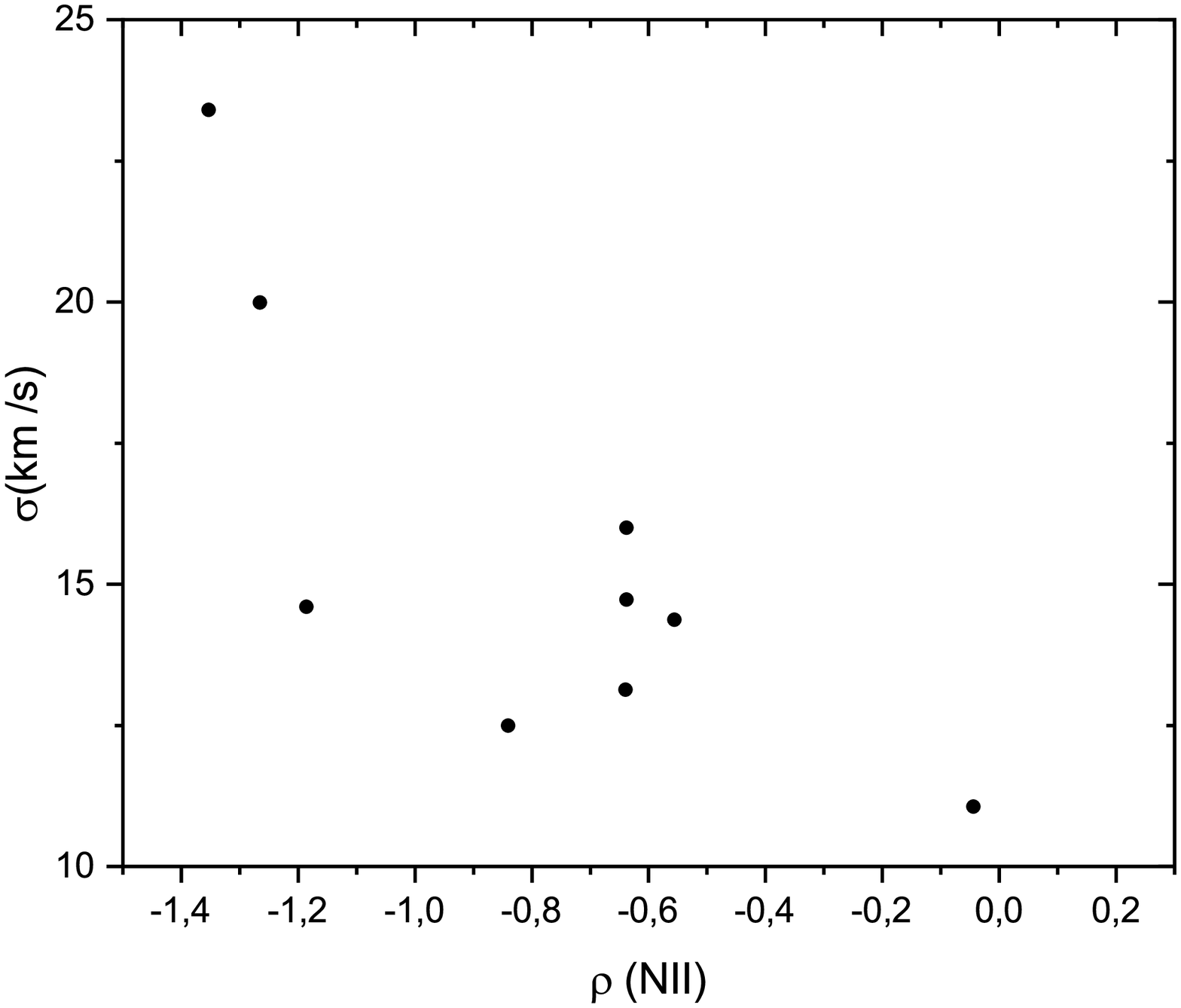}
\includegraphics[width=8 cm]{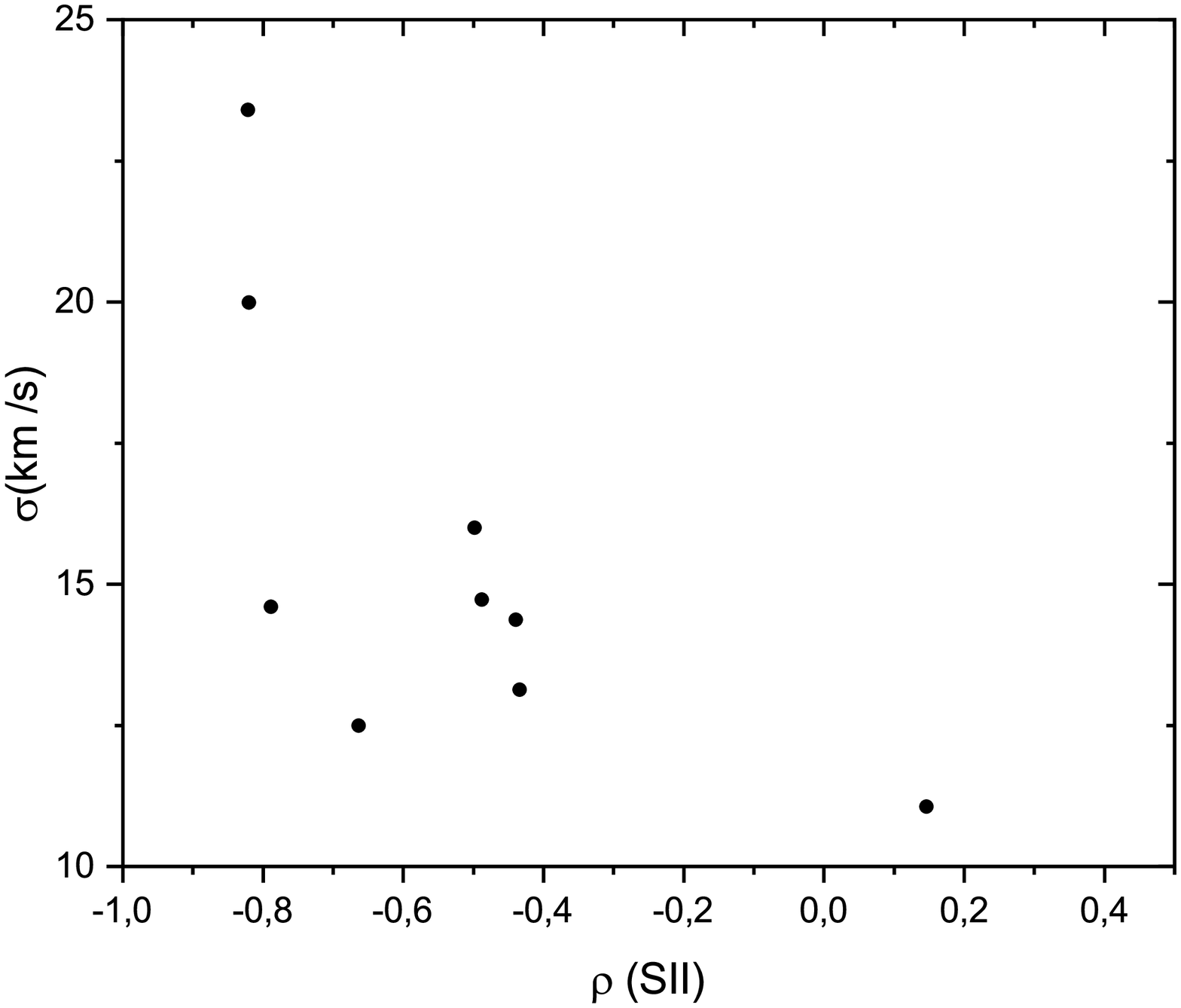} 
}
\caption{ The same as Fig. 2 for UGC 8508.}
\centering

\end{figure}

  
\section{Discussion}

We have investigated the BPT-$  \sigma$ and $  \rho$-$  \sigma$  diagrams for four nearby star-forming galaxies according to observations by the SAO RAS 6-meter telescope. The primary results of this study are as follows:
    
Mrk 370: The points with high velocity dispersion are interesting to study. These points indicate the presence of non-stellar mechanisms for the excitation of ionized gas, and the correlation between $  \rho$ and $  \sigma$ confirms the presence of shock waves in this galaxy.

NGC 4068: There is not an obvious correlation between $  \rho$ and $  \sigma$, but some points with high velocity dispersion should be studied.
   
UGC 8313: There is a correlation between $  \rho$ and $  \sigma$ for $  \sigma$ $ \leq $ 25 km/s. It is interesting to locate these points on the map to comment.

UGC  8508: Although it seems that there is not a correlation between $  \rho$ and $  \sigma$, our data is not enough to accurately comment on this issue. We did not detect a gas significantly ionized by shocks. The correlation between velocity dispersion and gas excitation is absent, so the gas excitation corresponds to the common situation in the galaxies with a relatively weak current star formation.

We are going to compare the characteristics of some galaxies to understand more about $  \rho$-$  \sigma$ correlation.
\section*{Acknowledgment}
This work is based on the observations of the 6-m telescope of the Special
Astrophysical Observatory of the Russian Academy of Sciences. The authors thank D. Oparin and others for their assistance in observations. 
Behjat Zarei Jalalabadi is also grateful for the financial support from the Ministry of Sciences, Researches and Technology of Islamic Republic of Iran.

\end{document}